%% file: nips_springer.tex
\documentclass{llncs}

\usepackage[utf8]{inputenc} 
\usepackage[T1]{fontenc}    
\usepackage{hyperref}       
\usepackage{url}            
\usepackage{booktabs}       
\usepackage{nicefrac}       
\usepackage{microtype}      
\usepackage[table]{xcolor}
\usepackage{amssymb,amsmath}
\usepackage{algorithm}
\usepackage{algpseudocode}
\usepackage{multicol}
\usepackage{tikz}

\DeclareMathOperator{\ReLU}{ReLU}
\DeclareMathOperator{\satReLU}{satReLU}
\DeclareMathOperator{\lReLU}{LeakyReLU}

\title{Automated and Formal Synthesis of Neural Barrier Certificates for Dynamical Models} 

\author{%
  Andrea~Peruffo\inst{1}, Daniele~Ahmed\inst{2}, Alessandro~Abate\inst{1} \\
}

\institute{Department of Computer Science,  
Department of Computer Science\\
University of Oxford, UK\\
  \email{\{name.surname\}@cs.ox.ac.uk}
\and 
Amazon Inc, London, UK
}

\begin{document}

\maketitle

\begin{abstract}
We introduce an automated, formal, counterexample-based 
approach to synthesise Barrier Certificates (BC) for the safety verification of continuous and hybrid dynamical models. 
The approach is underpinned by an inductive framework: 
this is structured as a sequential loop between a learner, 
which manipulates a candidate BC structured as a neural network,  
and a sound verifier, 
which either certifies 
the candidate's validity or generates counter-examples to further guide the learner. 
We compare the approach against state-of-the-art techniques, 
over polynomial and non-polynomial dynamical models: 
the outcomes show that we can synthesise sound BCs up to two orders of magnitude faster, 
with in particular a stark speedup on the verification engine (up to five orders less),   
whilst needing a far smaller data set (up to three orders less) for the learning part. 
Beyond improvements over the state of the art, 
we further challenge the new 
approach on a hybrid dynamical model and on larger-dimensional models, 
and showcase the numerical robustness of our algorithms and codebase. 
\end{abstract}

\section{Introduction}
\label{sec:intro}

Barrier Certificates (BC) are an effective and powerful 
technique to prove safety properties on models of continuous and hybrid dynamical systems~\cite{prajna2004safety,prajna2007framework}.   
%
%
Whenever found, 
a BC partitions the state space of the model into two parts, 
ensuring that all trajectories starting from a given initial set, located within one side of the BC, cannot reach a given set of states (deemed to be unsafe), located on the other side. 
Thus a successful synthesis of a BC (which is in general not a unique object) represents a formal proof of safety for the dynamical model.  
BC find various applications spanning robotics, multi-agent systems, and biology \cite{borrmann2015control,wang2017safety}.  

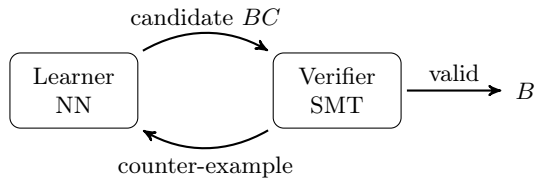
\begin{figure}
\centering 
\input{Figures/cegis}
\caption{Schematic representation of the CEGIS loop.}
\label{fig:cegis-loop}
\end{figure}

This work addresses the safety of dynamical systems modelled in general by non-linear ordinary differential equations (ODE), 
and presents a novel method for the automated and formal synthesis of BC.   
The approach leverages 
Satisfiability Modulo Theory (SMT) and inductive reasoning (CEGIS, Figure \ref{fig:cegis-loop}, introduced later),   
to guarantee the correctness of the automated synthesis procedure: 
this rules out both algorithmic and numerical errors related to BC synthesis \cite{dai2017barrier}. 

\paragraph{\textbf{Background and Related Work}}
A few techniques have been developed to synthesise BC.   
For polynomial models, sum-of-squares (SOS) and semi-definite programming relaxations~\cite{kong2013exponential,legat2020sum,sloth2012compositional}   
convert the BC synthesis problem into constraints expressed as linear or bilinear matrix inequalities: 
these are numerically solved as a convex optimisation problem, however unsoundly.  
To increase scalability and to enhance expressiveness, numerous barrier formats have been considered:  
BC based on exponential conditions are presented in \cite{kong2013exponential};  
BC based on Darboux polynomials are outlined in \cite{zeng2016darboux};  
\cite{sogokon2018vector} newly introduces a multi-dimensional generalisation of BC, thus broadening their scope and applicability. 
BC can also be used to verify safety of uncertain (e.g. parametric) models \cite{prajna2006barrier}. 
Let us remark that SOS approaches are typically \textit{unsound}, 
namely they rely on iterative and numerical methods to synthesise the BC. 
\cite{dai2017barrier} a-posteriori verifies SOS candidates via CAD techniques \cite{sk16}. 

%
Model \emph{invariants} (namely, regions that provably contain model trajectories, such as \textit{basins of attractions} \cite{NonLinSys:SS99}) can be employed as BC, 
though their synthesis is less general, as it does not comprise an unsafe set and tacitly presupposes the initial set to be ``well placed'' within the state space (that is, within the aforementioned basin): 
\cite{platzer2008computing} introduces a fixpoint algorithm to find algebraic-differential invariants for hybrid models; 
invariants can be characterised analytically \cite{ATS09} or synthesised computationally \cite{CASK20}. 
Invariants can be alternatively studied by \emph{Lyapunov theory} \cite{ahmed20automated}, 
which provides \emph{stability} guarantees for dynamical models, 
and thus can characterise invariants (and barriers) as side products:  
however this again requires that initial conditions are positioned around stable equilibria, 
and does not explicitly encompass unsafe sets in the synthesis. 
Whilst Lyapunov theory is classically approached either analytically (explicit synthesis) or numerically (with unsound techniques),  
an approach that is relevant for the results of this work looks at automated and sound Lyapunov function synthesis:   
in \cite{sankaranarayanan2013lyapunov} Lyapunov functions are soundly found within parametric templates,  
by constructing a system of linear inequality constraints over unknown coefficients. 
\cite{RS15,RS16,RS18} employ a counterexample-based approach to synthesise control Lyapunov functions, which inspires this work,  
using a combination of SMT solvers and convex optimisation engines: 
however unlike this work, SMT solvers are never used for verification, 
which is instead handled by solving optimisation problems that are numerically unsound. 
Let us emphasise again that the BC synthesis problem, as studied in this work, cannot in general be reduced to a problem of Lyapunov stability analysis, 
and is indeed more general.  


\paragraph{\textbf{Core approach} }
We introduce a method that efficiently exploits machine learning, 
whilst guaranteeing formal proofs of correctness via SMT. 
We leverage a CounterExample-Guided Inductive Synthesis (CEGIS) procedure~\cite{solar2006combinatorial}, 
which is structured as an inductive loop between a \emph{Learner} and a \emph{Verifier} (cf. Fig. \ref{fig:cegis-loop}).  
A learner numerically (and unsoundly) trains a neural network (NN) to fit over a finite set of samples the requirements for a BC, which are expressed through a loss function; 
then a verifier either formally proves the validity of the BC or provides (a) counter-example(s) through an SMT solver: the counter-examples indicate where the barrier conditions are violated, and are passed back to the learner for further training.  
%
%
This synthesis method for neural BC is formally sound and fully automated, 
and thanks to its specific new features, is shown to be much faster and to clearly require less data than state-of-the-art results. 

\paragraph{\textbf{Contributions beyond the State of the Art} }
Cognate work \cite{zhao2020barrier} presents a method to compute BC using neural networks and to verify their correctness a-posteriori: 
as such, it does not generate counter-examples within an inductive loop, as in this work. 
\cite{zhao2020barrier} considers large sample sets that are randomly divided into batches and fed to a feed-forward NN;  
the verification at the end of the (rather long) training either validates the candidate, or invalidates it and the training starts anew on the same dataset. 
In Section \ref{sec:experiments} the method in \cite{zhao2020barrier} is shown to be slower (both in the training and in the verification), 
and to require more data than the CEGIS-based approach of this work, 
which furthermore introduces numerous bespoke optimisations, as outlined in  Section \ref{sec:cegis-loop}:   
our CEGIS-based technique exploits fast learning, verification simplified by the candidates passed by the Learner, 
and an enhanced communication between Learner and Verifier.  
Our approach further showcases numerical robustness and scalability features.  

Related to the work on BC is the synthesis of Lyapunov functions, mentioned above.  
The construction of {\em Lyapunov Neural Networks} (LNNs) has been studied with approaches based on simulations and numerical optimisation, 
which are in general unsound \cite{richards2018lyapunov}. 
Formal methods for Lyapunov synthesis are introduced in \cite{ahmed20automated}, 
together with a  counterexample-based approach using polynomial candidates. 
The work is later extended in \cite{abate2020automated}, which employs NN as candidates over polynomial dynamical models.  
The generation of control Lyapunov functions using counterexample-based NN is similarly considered in \cite{ChangRG19}, 
however this is done by means of differing architectural details and with a different SMT solver, and does not extend to BC synthesis.  
Beyond the work in \cite{ahmed20automated}, 
this contribution is not limited to a specific polynomial template, 
since it supports more general mixtures of polynomial functions obtained through the NN structure, 
as well as the canonical tanh, sigmoid, ReLU activations (we provide one example of BC using tanh activations).  
Compared to \cite{ahmed20automated}, where we use LP programming to synthesise Lyapunov functions, we now: 
$a)$ use a template-free procedure, thanks to the integration of NNs - these are needed since template-based SOS-programming approaches are not sufficient to provide BCs for several of the presented benchmarks (see Section \ref{sec:experiments} and \cite{zhao2020barrier});
$b)$ we provide an enhanced loss function (naturally absent from \cite{ahmed20automated}), enriched counterexample generation, prioritised check of the verification constraints, and
$c)$ we handle hybrid models, for which using formal methods represents a novelty in the literature. 
Finally, beyond \cite{ahmed20automated} the new approach is endowed with numerical robustness features.  

SOS programming solutions \cite{kong2013exponential,legat2020sum,sloth2012compositional} are not quite comparable to this work. 
Foremostly, they are not sound, i.e. do not offer a formal guarantee of numerical and algorithmic correctness. 
(The exception is \cite{dai2017barrier}, which verifies SOS candidates a-posteriori by means of CAD \cite{sk16} techniques that are known not to scale well.)  
Furthermore, they can be hardly embedded within a CEGIS loop - we experimentally show that SOS candidates are handled with difficulty by SMT solvers. 
Finally, they hardly cope with the experiments we have considered, as already observed in \cite{zhao2020barrier}. 
We instead use SMT solvers (Z3 \cite{de2008z3} and dReal \cite{gao2013dreal}) within CEGIS to provide sound outcomes based on NN candidates, 
proffering a new approach that synthesises and formally verifies candidate BCs altogether, with minimum effort from the user.

\paragraph{\textbf{Organisation} }
The remainder of the paper is organised as follows:
Section \ref{sec:barrier-certificates} presents preliminary notions on BCs and outlines the problem. 
Section \ref{sec:cegis-loop} describes the approach to synthesise barrier certificates: 
Sec. \ref{subsec:learner} introduces the neural training technique, whereas Sec. \ref{subsec:verifier} illustrates SMT verification in relation to our framework.  
Section \ref{sec:experiments} presents case studies,  
Section \ref{sec:conlusion} delineates future work. 



\section{Safety Analysis with Barrier Certificates}
\label{sec:barrier-certificates}

We address the safety verification of continuous-time dynamical models by designing barrier certificates (BC) over the continuous state space $X$ 
of the model. 
We consider $n$-dimensional dynamical models described by 
\begin{equation}
\dot{x}(t) = \frac{dx}{dt} = f(x),  \quad  x(0) = x_0 \in X_0 \subset X, 
\label{eq:sys-dynamics}
\end{equation}
where $f: X \rightarrow \mathbb{R}^n$ is a continuous vector field, $X \subseteq \mathbb{R}^n$ is an open set  defining the state space of the system, and $X_0$ represents the set of initial states. 
Given model \eqref{eq:sys-dynamics} and an unsafe set $X_u\subset X$, 
the safety verification problem concerns checking whether or not all trajectories of the model originating from $X_0$ reach the unsafe region $X_u$. 
BC offer a sufficient condition asserting the safety of the model, namely when no trajectory enters the unsafe region. 

\begin{definition}
The Lie derivative of a continuously differentiable scalar function $B: X \rightarrow \mathbb{R}$, 
with respect to a vector field $f$, 
is defined as follows 
\begin{equation}
\label{eq:lie}
\dot{B}(x) = \nabla B(x) \cdot f(x) = 
\sum_{i=1}^n \frac{\partial B}{\partial x_i} \frac{d x_i}{dt} = 
\sum_{i=1}^n \frac{\partial B}{\partial x_i} f_i(x). 
\end{equation}
Intuitively, this derivative denotes the rate of change of function $B$ along the model trajectories.
\end{definition}

\begin{proposition}[Barrier Certificate for Safety Verification, \cite{prajna2004safety}]
Let the model in \eqref{eq:sys-dynamics} and the sets $X$, $X_0$ and $X_u$ be given. 
Suppose there exists a function $B: X \rightarrow \mathbb{R}$ that is differentiable with respect to its argument and satisfies the following conditions:
\begin{equation}
B(x) \leq 0 \ \forall x \in X_0, 
\quad
B(x) > 0 \ \forall x \in X_u, 
\quad
\dot{B}(x) \leq 0 \ \forall x \in X \text{ s.t. } B(x) = 0, 
\label{eq:barrier-conditions}
\end{equation}
then the safety of the model is guaranteed. 
That is, there is no trajectory of the model contained in $X$, 
starting from the initial set $X_0$, that ever enters set $X_u$. 
\end{proposition}
Consider a trajectory $x(t)$ starting in $x_0 \in X_0$ and the evolution of $B(x(t))$ along this trajectory. 
Whilst the first of the three conditions guarantees that $B(x_0) < 0$,  the last condition asserts that the value of $B(x(t))$ along a trajectory $x(t)$ cannot become positive. 
Hence such a trajectory $x(t)$ cannot enter the set $X_u$, where $B(x) > 0$ (second condition), 
thus ensuring the safety of the model.


\section{Synthesis of Neural Barrier Certificates via Learning and Verification}  
\label{sec:cegis-loop}

We introduce an automated and formal approach for the construction of barrier certificates (BC) that are expressed as feed-forward neural networks (NN).  
The procedure leverages CEGIS (see Fig. \ref{fig:cegis-loop})~\cite{solar2006combinatorial}, 
an automated and sound procedure for solving second-order logic synthesis problems, 
which comprises two interacting parts. 
The first component is a \emph{Learner}, which provides candidate BC functions by training a NN over a finite set of sample inputs. 
The candidate is passed to the second component, a \emph{Verifier}, which acts as an oracle: 
either it proves that the solution is valid, 
or it finds one (or more) instance (called a counter-example) 
where the candidate BC does not comply with required conditions.  
The verifier consists of an SMT solver~\cite{sk16}, 
namely an algorithmic decision procedure that extends Boolean SAT problems to richer, more expressive theories, 
such as non-linear arithmetics.  

More precisely, 
the learner trains a NN composed of $n$ of input neurons (this matches the dimension of the model $f$), $k$ hidden layers, and one output neuron (recall that $B(x)$ is a scalar function):
this NN candidate $B$ is required to closely match the conditions in Eq. \eqref{eq:barrier-conditions} over a discrete set of samples $S$,
which is initialised randomly. 
The verifier checks whether the candidate $B$ violates any of the conditions in Eq. \eqref{eq:barrier-conditions} over the entire set $X$ and, if so, produces one (or more, as in this work) counter-examples $c$. 
We add $c$ to the samples set $S$ as the loop restarts, hence forcing the NN to be trained \emph{also} over the generated counter-examples $c$. 
This loop repeats until the SMT verifier proves that no counter-examples exist or until a timeout is reached. 
CEGIS offers a scalable and flexible alternative for BC synthesis: 
on the one hand, the learner does not require soundness, and ensures a rapid synthesis exploiting the training of NN architectures;  
on the other, the algorithm is {\it sound}, i.e. a valid output from the SMT-based verifier is provably correct;  
of course we cannot claim any {\it completeness}, since CEGIS might in general not terminate with a solution because it operates over a continuous model and its domain $X$.  

The performance of the CEGIS algorithm in practice hinges on the effective exchange of information between the learner and the verifier \cite{bcADKKP18}. 
A core contribution of this work is to tailor the CEGIS architecture to the problem of BC synthesis:  
we devise several improvements to NN training, 
such as a bespoke loss function and a multi-layer NN architecture that ensures robustness and outputs a function that is tailored to the verification engine, 
together with informative counter-examples generation by the SMT verifier that is adapted to the candidate BC and the underlying dynamical model. 
These tailored architectural details generate in practice a rapid, efficient, and robust CEGIS loop, 
which is shown in this work to clearly outperform state-of-the-art methods.

\subsection{Training of the Barrier Neural Network}
\label{subsec:learner}

The learner instantiates the candidate BC using the hyper-parameters $k$ and $h$ (depth and width of the NN), 
trains it over the $N$ samples in the set $S$, 
and later refines its training whenever the verifier adds counter-examples to the set $S$.  
The class of candidate BC comprises multi-layered, feed-forward NN with 
\emph{polynomial} activation functions. 
Unlike most learning applications, 
the choice of polynomial activations comes from the need for interpretable outputs from the NN,  
whose analytical expression must be readily processed by the verifier. 
\footnote{Whilst this work emphasises the novel use of polynomial activation functions, 
we remark that alternatives are possible: in particular, as displayed in one case study, 
$\tanh$ are well-suited to our objective, 
inducing universal function approximators.}   
  The order $\gamma$ of the polynomial activations is a hyper-parameter fed at the start of the CEGIS procedure: 
 we split the $i$-th hidden layer into $\gamma$ portions and apply polynomial activations of order $j$ to the neurons of the $j$-th portion, 
  as shown next. 
\begin{example}[Polynomial Activations]
Assume a NN composed of an input $x$, 5 hidden neurons and 1 activation-free output, with $\gamma$-th order polynomial activation, $\gamma = 5$. 
We split the hidden layer in $\gamma$ sub-vectors, each containing one neuron. 
The hidden layer after the activation results in
\begin{equation*}
z =  \begin{bmatrix}
 W_1^{(1)} x + b_1 & (W_2^{(1)} x + b_2)^2 & (W_3^{(1)} x + b_3)^3 & (W_4^{(1)} x + b_4)^4 & (W_5^{(1)} x + b_5)^5
\end{bmatrix} ^T,
\end{equation*}
where the $W_i^{(1)}$ are the $i$-th row of the first-layer weight matrix, 
and the $b_i$ form the bias vector.  \qed
\end{example}  
The learning process updates the NN parameters
 to improve the satisfaction of the BC conditions in \eqref{eq:barrier-conditions}:  
 $B(x) \leq 0$ for $x \in X_0$, $B(x) > 0$ for $x \in X_u$, 
 and a negative Lie derivative $\dot{B}$ (eq. \eqref{eq:lie}) over the set implicitly defined by $B(x) = 0$. 
The training minimises a loss comprising thee terms, namely  
\begin{multline}
L = L_0 + L_u + L_d = 
\frac{1}{N} \sum_{i=1}^N 
\Big(\max_{s_i \in X_0}\{\tau_0, B(s_i)\} 
+ \max_{s_i \in X_u}\{\tau_u, -B(s_i)\} \\
+ \max_{s_i:B(s_i)=0}\{\tau_d, \dot{B}(s_i)\}\Big),
\label{eq:loss-function-1}
\end{multline} 
where $s_i$, $i = 1, \dots, N$ are the samples taken from the set $S$. 
The constants $\tau_0$, $\tau_u$, $\tau_d$ are offsets, added to improve the numerical stability of the training. 
Notably, 
$B(x) = 0$ is a measure-zero set, thus it is highly unlikely that a single sample $s$ will satisfy $B(s) = 0$. 
We then relax this last condition and consider a belt $\mathcal{B}$ around $B(s) = 0$, 
namely $\mathcal{B} = |B(x)| \leq \beta$, 
which depends on the hyper-parameter $\beta$. 
%
%
Note that we must use continuously differentiable activations throughout,  
as we require the existence of Lie derivatives (cf. Eq. \eqref{eq:lie}), 
and thus cannot leverage simple ReLUs. 

\paragraph{\textbf{Enhanced Loss Functions}}
%
The loss function in Eq. \eqref{eq:loss-function-1} experimentally yields possible drawbacks, 
which suggest a few ameliorations.  
Terms $L_0$ and $L_u$ solely penalise samples with incorrect value of $B(x)$ without further providing a reward for samples with a correct value. 
The NN thus stops learning when the samples return correct values of $B(x)$ without further increasing the positivity of $B$ over $X_u$ or the negativity over $X_0$.
As such, the training often returns a candidate $B(x)$ with values just below $\tau_0$ in $X_0$ or above $\tau_u$ in $X_u$. 
These candidates are easily falsified, thus potentially leading to a large number of CEGIS iterations. 

We improve the learning by adopting a (saturated) {\it Leaky} ReLU, hence rewarding samples that evaluate to a correct value of $B(x)$. 
Noting that 
\begin{equation}
\lReLU(\alpha, x) = \ReLU(x) - \alpha \ReLU(-x),
\end{equation}
where $\alpha$ is a small positive constant, we rewrite  term $L_0$ as
\begin{equation}
L_0 = 
\frac{1}{N} \sum_{s_i \in X_0} 
\ReLU(B(s_i) - \tau_0) - \alpha \cdot \satReLU(- B(s_i) + \tau_0),
\end{equation}
where $\satReLU$ is the saturated $\ReLU$ function. Term $L_u$ is similarly modified. 
In view of the composite nature of our training objective, 
incorrect samples account for the major contribution to the loss function, 
leading the NN to correct those first. 
At a second stage, the network finds a direction of improvement by following the {\it leaky} portion of the loss function. 
This is saturated to prevent the training from following only one of these directions, 
without improving the other loss terms. 

Another possible drawback of the loss function in \eqref{eq:loss-function-1} derives from the term $L_d$: 
it solely accounts for a penalisation of the sample points within $\mathcal{B}$. 
To quickly and myopically improve the loss function,  
the training can generate a candidate BC for which no samples are within $\mathcal{B}$ - 
we experimentally find that this behaviour persists, regardless of the value of $\beta$.  
Similarly to $L_0$ and $L_u$, we reward the points within a belt fulfilling the BC condition:  
namely, we solely apply the $\satReLU$ function to reward samples $s$ with a negative $\dot{B}(s)$, 
whilst not penalising values $\dot{B}(s) \geq 0$. 
The training is driven to include more samples in $\mathcal{B}$, 
guiding towards a negative $\dot{B}(s)$, and finally enhancing learning. 
The expression of $L_d$ results in
\begin{equation}
L_d = - \frac{1}{N} \sum_{s \in \mathcal{B}}
\satReLU(- \dot{B}(s) + \tau_d).
\label{eq:Ld-improved}
\end{equation}
Finally, we choose an asymmetric belt $\mathcal{B} =  - \beta_1 \leq B(s) \leq \beta_2$, with $\beta_2 > \beta_1 > 0$ to both ensure a wider sample set and a stronger safety certificate. 

\paragraph{\textbf{Multi-layer Networks}}
Polynomial activation functions generate interpretable barrier certificates with analytical expressions that are readily verifiable by an SMT solver. 
However, when considering polynomial networks, the use of multi-layer architectures quickly increases the order of the barrier function: 
a $k$-layer network with $\gamma$-th order activations returns a polynomial of $k \gamma$ degree. 
We have experienced that deep NN provide numerical robustness to our method, 
although the verification complexity increases with the order of the polynomial activation functions used and with the depth of the NN.
As a consequence, our procedure leverages a deep architecture whilst maintaining a low-order polynomial by interchanging linear and polynomial activations over adjacent layers.    
We have observed that the use of linear activations, particularly in the output layer, positively affect the training:  
they providing robustness that is needed to the synthesis of BC (see Experimental results), 
without increasing the order of the network with new polynomial terms. 

\paragraph{\textbf{Learning in Separate Batches}}
The structure of the conditions in \eqref{eq:barrier-conditions} and the learning loss in \eqref{eq:loss-function-1} naturally suggests a separate, parallel approach to training.  
We then split the dataset $S$ into three batches $S_0$, $S_u$ and $S_x$, 
each including samples belonging to $X_0$, $X_u$ and $X$, 
respectively.  
For training, we compute the loss function in a parallel fashion. 
Similarly, for the verifier, generated counter-examples are added to the relevant batch.


\subsection{Certification of the Barrier Neural Network, or Falsification via Counter-examples}
\label{subsec:verifier}

Every candidate BC function $B(x)$ which the learner generates requires to be certified by the verifier. 
Equivalently, 
in practice the SMT-based verifier aims at finding states that violate the barrier conditions in \eqref{eq:barrier-conditions} over the continuous domain $X$.  
To this end, 
we express the {\it negation} of such requirements, and formulate a nonlinear constrained problem over real numbers, as  
\begin{equation}
(x \in X_0 \wedge B(x) > 0) \vee
(x \in X_u \wedge B(x) \leq 0) \vee
(B(x) = 0 \wedge \dot{B}(x) > 0).
\label{eq:verifier-conditions}
\end{equation}
The verifier searches for solutions of the constraints in Eq. \eqref{eq:verifier-conditions},   
which in general requires manipulating non-convex 
functions. 
%
On the one hand, the soundness of our CEGIS procedure heavily relies on the correctness of SMT solving: 
an SMT solver never fails to assert the absence of solutions for \eqref{eq:verifier-conditions}. 
%
As a result, when it states that formula \eqref{eq:verifier-conditions} is unsatisfiable, i.e. returns {\tt unsat},  
$B(x)$ is formally guaranteed to fulfil the BC conditions in Eq. \eqref{eq:barrier-conditions}. 
On the other hand, the CEGIS algorithm offers flexibility in the choice of the verifier, 
hence we implement and discuss two SMT solvers: dReal \cite{gao2013dreal} and Z3 \cite{de2008z3}.   
dReal is a $\delta$-complete solver,  
namely the \texttt{unsat} decision is correct \cite{gao2012delta}, 
whereas when a solution for \eqref{eq:verifier-conditions} is found, this comes with a $\delta$-error bound.  
The value of $\delta$ characterises the procedure precision. 
In our setting, it is acceptable to return spurious counter-examples: 
indeed, these are then used as additional samples and do not invalidate the sound outcomes of the procedure, but rather help synthesising a more robust barrier candidate. 
dReal is capable of handling non-polynomial terms, 
such as as exponentials or trigonometric vector fields $f$ for some of the models considered in Section \ref{sec:experiments}. 
Z3 is a powerful, sound and complete SMT solver, 
namely its conclusions are provably correct both when it determines the validity of a BC candidate and when it provides counter-examples. 
The shortcoming of Z3 is that it is unable to handle non-polynomial formulae.  


\paragraph{\textbf{Prioritisation and Relaxation of Constraints} }
The effectiveness of the CEGIS framework is underpinned by rapid exchanges between the learner and the verifier, 
as well as by quick NN training and SMT verification procedures.  
We have experienced that the bottleneck resides in the handling of the constraint $\eta_d = (B(x)=0 \wedge \dot{B}(x) > 0)$ by the SMT solver,   
since the formula contains the high-order expression $\dot{B}(x)$ and because it is defined over the thin region of the state space implicitly characterised by $B(x)=0$.  
As a consequence, 
we have prioritised constraints $\eta_0 = (x \in X_0 \wedge B(x) > 0)$ and $\eta_u = (x \in X_u \wedge B(x) \leq 0)$: 
that is, if either clauses is satisfied, i.e. a counter-example is found for at least one of them, 
the verifier omits testing $\eta_d$ whilst the obtained counter-examples are passed to the learner.  
The constraint $\eta_d$ is thus checked solely if $\eta_0$ and $\eta_u$ are both deemed to be {\tt unsat}. 
Whenever this occurs, and the verification of $\eta_d$ times out, 
the solver searches for a solution of a relaxed constraint $(|B(x)|<\tau_v \wedge \dot{B}(x) > 0)$, 
similarly to the improved learning conditions discussed in Eq. \eqref{eq:Ld-improved}.   
Whilst this constraint is arguably easier to solve in general, it may generate spurious counter-examples, 
namely a sample $\bar{x}$  that satisfy the relaxed constraint, but such that $B(\bar{x})\neq0$.  
The generation of these samples does not contradict the soundness of the procedure, 
and indeed are shown to improve the robustness of the next candidate BC -- this of course comes with the cost of increasing the number of CEGIS iterations.

\paragraph{\textbf{Increased Information from Counter-examples} }
The verification task encompasses an SMT solver attempting to generate a counter-example, 
namely a (single) instance satisfying Eq. \eqref{eq:verifier-conditions}. 
However, a lone sample might not always provide insightful information for the learner to process. 
Na\"ively asking the SMT solver to generate more than one counter-example can be in general expensive, as it is done sequentially. 
Specifically, the verifier solves Eq. \eqref{eq:verifier-conditions} to find a first counter-example $\bar{x}$; 
then, to find any additional sample, we include the statement $(x \neq \bar{x})$ and solve again for the resulting formula.  
We are interested in finding numerous points invalidating the BC conditions and feed them to the learner as a batch,  
or in increasing the information generated by the verifier by finding a sample that maximises the violation of the BC conditions.  
To this end,  
firstly we randomly generate a {\it cloud} of points around the generated counter-example: 
in view of the continuity of the candidate function $B$, samples around a counter-example are also likely to invalidate the BC conditions.  
Secondly, for all points in the cloud, we compute the gradient of $B$ (or of $\dot{B}$) and follow the direction that maximises the violation of the BC constraints. 
As such, we follow the $B$ (resp. $\dot{B}$) maximisation when considering $x \in X_0$ ($x \ s.t. \ |B(x)| < \tau_v$), 
and viceversa when $x \in X_u$.  
This gradient computation is extremely fast as it exploits the neural architecture, 
and it provides more informative samples for further use by the learner. 
%

\input{Figures/algorithm}


\section{Case Studies and Experimental Results}
\label{sec:experiments} 

We have implemented the discussed new procedure using the PyTorch library. 
All experiments are performed on a laptop workstation with 8 GB RAM, running on Ubuntu 18.04.   
We demonstrate that the proposed method finds provably correct BCs on benchmarks from literature comprising both polynomial and non-polynomial dynamics: 
we compare our approach against the work 
\cite{zhao2020barrier}, as this is the only work on sound synthesis of BCs with NNs to the best of our knowledge, 
and against the well-know SOS optimisation software SOSTOOLS \cite{sostools}. 
Beyond the benchmarks proposed in \cite{zhao2020barrier}, 
we newly tackle a hybrid model 
as well as larger, (up to) 8-dimensional models, which push the boundaries of the verification engine and display a significant extension to the state of the art. 
To confirm the flexibility of our architecture in accepting different SMT solvers,  
we verify generated candidate BC using dReal in the first four benchmarks, whereas we study the last four using Z3.    
In all the examples, we use a learning rate of 0.1 for the NN and the loss function described in Section \ref{subsec:learner} with $\alpha = 10^{-4}$, $\tau_0 = \tau_u = \tau_d = 0.1$. 
The region in Eq. \eqref{eq:Ld-improved} is limited by $\beta_1 = 0.1$ whilst $\beta_2 = \infty$. 
We set a verification parameter $\tau_v = 0.05$ (cf. Sec. \ref{subsec:verifier}), a timeout (later denoted as OOT) of 60 seconds and the 
precision for dReal to $\delta = 10^{-6}$. 
Table \ref{tab:results} summarises the outcomes.  

For the first four benchmarks, we compare our procedure, denoted as CEGIS, with the repeated results from \cite{zhao2020barrier}, 
which however does not handle the hybrid model in the fifth benchmark.  
We have run the algorithm in \cite{zhao2020barrier} and reported the cumulative synthesis time under the `Learn' column.  
However the verification (we emphasise this is done only once, a-posteriori, in \cite{zhao2020barrier}) is not included in the repeatability package, 
hence we report the results from \cite{zhao2020barrier}, which are generated with much more powerful hardware than the one used for this work.  
Due to this issue of lack of repeatability, we have not run \cite{zhao2020barrier} on the larger-dimensional ODE models.
Compared to \cite{zhao2020barrier}, the outcomes suggest that we obtain \emph{much faster} synthesis and verification times (up to five orders of magnitude), 
whilst requiring up to only 0.02\% (see Obstacle Avoidance Problem) of the training data: 
\cite{zhao2020barrier} performs a uniform sampling of the space $X$, hence suffers especially in the 3-d case, 
where the learning runs \emph{two orders of magnitude} faster.  
Evidently 
this gap in performance derives from the different synthesis procedure:  
it appears to be more advantageous to employ a much smaller, randomly sampled, initial dataset that is progressively augmented with counter-examples,  
rather than 
to uniformly sample the state space to then train the neural network over a larger dataset. 

Next, we have implemented the SOS optimisation problems in \cite{dai2017barrier} within the software SOSTOOLS \cite{sostools} to generate barrier candidates, 
which are polynomials up to order 4 (this is the max order of the polynomial candidates generated by our Learner). 
In a few instances we ought to conservatively approximate the expression of $X_0$ or $X_u$ in order to encode them as SOS program - this makes their applicability less general. 
SOSTOOLS has successfully found BC candidates for five of the eight benchmarks, 
and they were generated consistently fast, in view of the convex structure of the underlying optimisation problem.  
However, recall that these techniques lack soundness (also due to numerical errors), which is instead a core asset of our approach. 
Consequently, we have passed them to the Z3 SMT solver, which should easily handle polynomial formulae:  
only one of them (`Hybrid Model') has been successfully verified; 
instead, the candidate for the `Polynomial Model' has been invalidated (namely Z3 has found a counter-example for it),   
whereas the verification of the remaining BC candidates has run out of time. 
For the latter instances, 
we have experienced that SOSTOOLS generally returns numerically ill-conditioned expressions, 
namely candidates with coefficients of rather different magnitude, with many decimal digits: 
even after rounding, 
expressions with this structure are known to be hardly handled by SMT solvers \cite{abate2020automated,ahmed20automated}, 
which results in long time needed to return an answer - this explains the experienced timeouts. 
These experiments suggest that the use of SOS programs within a CEGIS loop appears hardly attainable. 

Notice that all the case studies are solved with a \emph{small number} of iterations (up to 8) of the CEGIS loop:  
this feature, along with the limited runtimes, 
is promising towards tackling synthesis problems over more complex models. 
%

For the eight case studies,  
we report below the full expressions of the dynamics of the models,  
the spatial domain $X$ (as a set of constrains), 
the set of initial conditions $X_0 \subset X$, 
and the unsafe set $X_u \subset X$. 
We add a detailed analysis of the CEGIS iterations involved in the synthesis of the corresponding BCs.  

\begin{table}[h]
\centering
\begin{tabular}{c||c||c||c}
Benchmark & CEGIS (this work) & BC from \cite{zhao2020barrier} & SOS from \cite{sostools} 
\\ \hline
 \begin{tabular}{c}
 \\
  Darboux \\
  Exponential \\
  Obstacle \\
  Polynomial\\
  Hybrid  \\
  4-d ODE  \\
  6-d ODE  \\
  8-d ODE  \\
 \end{tabular}
 & 
 \begin{tabular}{cccc}
 {\footnotesize Learn} & {\footnotesize Verify} & {\footnotesize Samples} & {\footnotesize Iters} \\ \hline
27.2 & 0.04 & 0.5 & 1 \\
13.9 & 0.01 & 0.5 & 2 \\
19.6 & 0.01 & 0.5 & 1 \\
64.3 & 0.02 & 1 & 2  \\ 
8.97 & 15.51 & 0.5 & 8 \\
76.75 & 0.17 & 1 & 2 \\
74.92 & 1.30 & 1 & 2 \\
95.83 & 74.62 & 1 & 2 \\
 \end{tabular}
 & 
 \begin{tabular}{ccc}
 {\footnotesize Learn} & {\footnotesize Verify} & {\footnotesize Samples} \\ \hline
54.9 & 20.8 & 65 \\
234.0 & 11.3 & 65 \\
3165.3 & 1003.3 & 2097 \\
1731.0 & 635.3 & 65 \\
 -- & -- & -- \\
 -- & -- & -- \\
 -- & -- & -- \\
 -- & -- & -- \\
 \end{tabular}
& 
 \begin{tabular}{cc}
 {\footnotesize Synth} & {\footnotesize Verify} 
 \\ \hline
$\times$ & --  \\
$\times$ & --  \\
$\times$ & --  \\
8.10 & $\times$ \\
 12.30 & 0.11  \\
 12.90 & OOT  \\
 16.60 & OOT  \\
 26.10 & OOT  \\
 \end{tabular}
\end{tabular}
\caption{Outcomes of the case studies:  
Cumulative time for Learning and Verification steps are given in seconds; 
`Samples' indicates the size of input data for the Learner (in thousands); 
`Iters' is the number of iterations of the CEGIS loop (which is specific to our work); 
$\times$ indicates a synthesis or verification failure; OOT denotes a verification timeout. 
The Hybrid and the three ODE Models are newly introduced in this work.  
}
\label{tab:results}
\end{table}

\begin{figure}[h!]
\centering
\includegraphics[trim={1.75cm 0cm 1.75cm 1.75cm}, width=0.49\textwidth]{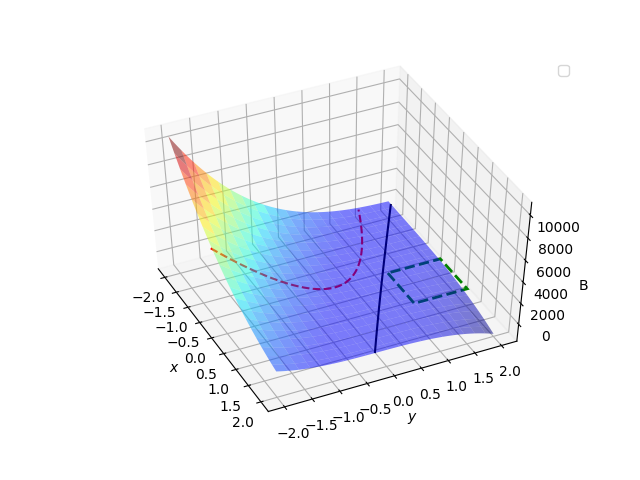}
\includegraphics[width=0.49\textwidth]{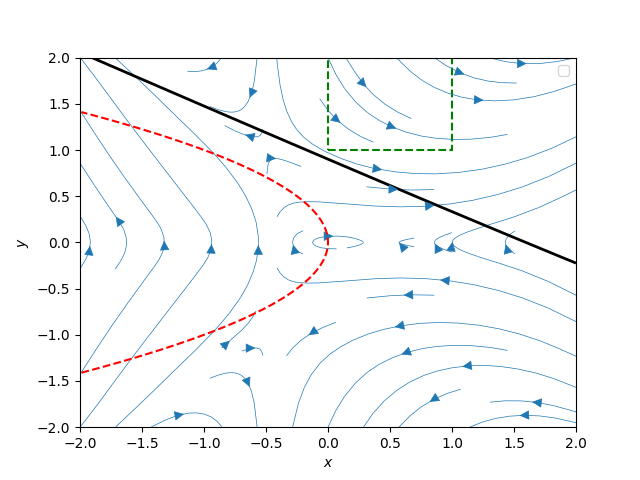}
\includegraphics[trim={1.75cm 0cm 1.75cm 1.75cm}, width=0.49\textwidth]{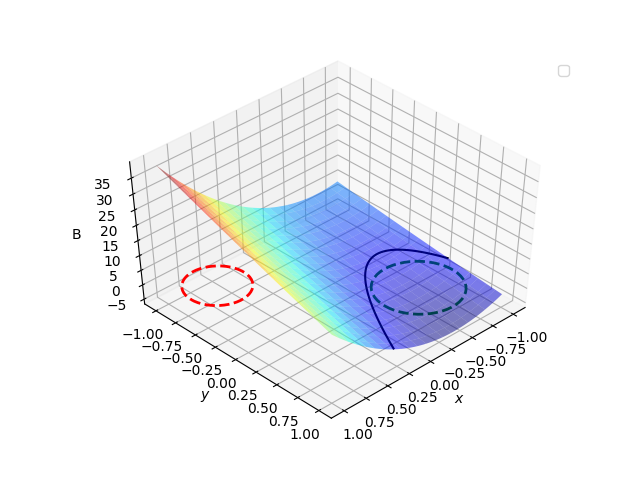}
\includegraphics[width=0.49\textwidth]{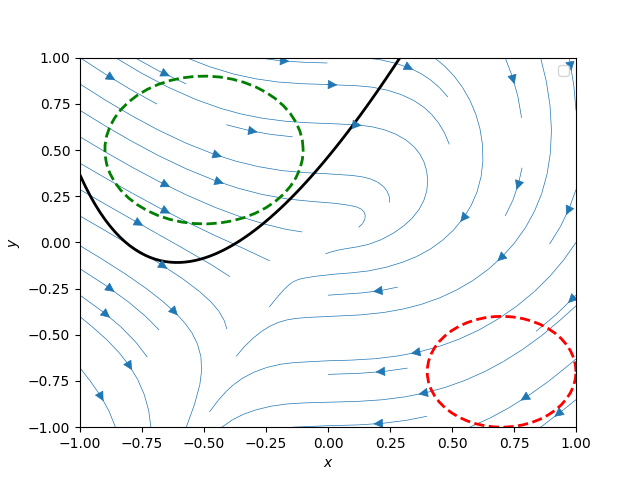}
\includegraphics[trim={1.75cm 1.75cm 1.75cm 1.75cm}, width=0.49\textwidth]{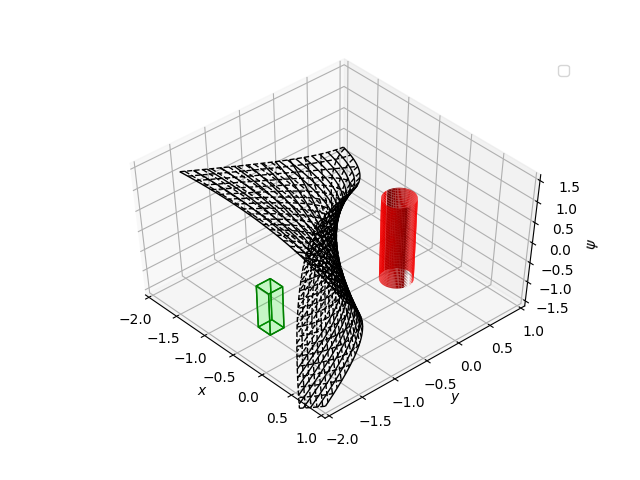}
\includegraphics[width=0.49\textwidth]{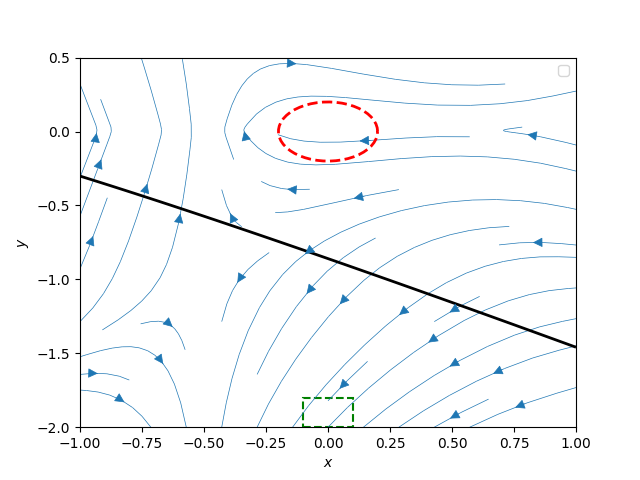}
\caption{
The BC for the Darboux (top left), Exponential (middle left), and Obstacle Avoidance (the 3D study, bottom left) models with corresponding vector fields (right column).  
Initial and unsafe sets are represented in green and red, respectively; the black line outlines the level curve $B(x)=0$. 
}
\label{fig:three-case-studies}
\end{figure}

\paragraph{\textbf{Darboux Model} }
This 2-dimensional model 
is approached using polynomial BCs. Its analytical expression is
\begin{equation*}
\begin{cases}
\dot{x} = y + 2xy, 
\\
\dot{y} = -x + 2x^2 - y^2,
\end{cases}
\quad 
\text{ with domains }
\quad
\begin{aligned}
& X = \{ -2 \leq x, y \leq 2 \}, 
\\
& X_0 =  \{ 0 \leq x \leq 1, 1 \leq y \leq 2 \}, 
\\
& X_u = \{ x+y^2 \leq 0 \}.
\end{aligned}
\end{equation*}
The work \cite{zeng2016darboux} reports that LMI-based methods fail to verify this model using polynomial templates of degree 6. 
Our approach generates the BC shown in Fig. \ref{fig:three-case-studies} (left) in less than 30 seconds, roughly half as much as in \cite{zhao2020barrier},  
and using only 500 initial samples vs more than 65000.
The initial and unsafe sets are depicted in green and red, respectively, whereas the level set $B(x) = 0$ is outlined in black.  
The BC is derived from a 3-layer architecture of 10 nodes each, with linear, polynomial with $\gamma=3$, and linear activations respectively.

\paragraph{\textbf{Exponential Model} }
This model from \cite{liu2015abstraction} 
shows that our approach extends to non-polynomial systems encompassing exponential and trigonometric functions: 
\begin{equation*}
\begin{cases}
\dot{x} = e^{-x} + y - 1, \\
\dot{y} = - \sin^2 x,
\end{cases}
\quad 
\text{ with domains }
\quad
\begin{aligned}
& X = \{-2 \leq x,y  \leq 2 \}, \\ 
& X_0 = \{(x+0.5)^2+(y - 0.5)^2 \leq 0.16 \}, \\ 
& X_u  = \{ (x - 0.7)^2+(y+0.7)^2 \leq 0.09 \}.
\end{aligned}
\end{equation*}
Our algorithm provides a valid BC in 14 seconds, 
around 5\% of the results in \cite{zhao2020barrier}, again using solely 500 initial samples. 
The BC, depicted in Fig.\ref{fig:three-case-studies} (centre), results from a single-layer neural architecture of 10 nodes, with polynomial activation function with $\gamma=3$.

\paragraph{\textbf{Obstacle Avoidance Problem}}
This 3-dimensional model, originally presented in \cite{barry2012safety}, describes a robotic application: 
the control of the angular velocity of a two-dimensional airplane, aimed at avoiding a still obstacle. The details are 
\begin{equation*}
\begin{cases}
\dot{x} = v \sin \varphi, \\
\dot{y} = v \cos \varphi, \\
\dot{\varphi} = u, \quad \textrm{ where } \quad u = - \sin \varphi + 3 \cdot \dfrac{x \sin \varphi + y \cos \varphi}{0.5 + x^2 + y^2}, \text{ with domains }
\end{cases} 
\end{equation*}
\vspace{-0.2in}
\begin{align*}
& X = \{ -2 \leq x,y \leq 2,-\nicefrac{\pi}{2}< \varphi < \nicefrac{\pi}{2} \}, \\
& X_0 = \{ -0.1 \leq x \leq 0.1, - 2 \leq y \leq -1.8,-\nicefrac{\pi}{6}< \varphi< \nicefrac{\pi}{6} \}, \\
& X_u = \{ x^2+ y^2  \leq 0.04 \}.  
\end{align*}
The BC is obtained from a single-layer NN comprising 10 neurons, using polynomial activations with $\gamma=2$. 
Fig. \ref{fig:three-case-studies} (right) plots the vector field on the plane $z=0$. 
Our procedure takes 0.6\% of the computational time in \cite{zhao2020barrier}, 
providing a valid BC with 1 iteration starting from an initial dataset of 500 samples.

\begin{figure}
\centering
\includegraphics[trim={1.5cm 1.5cm 1.5cm 1.5cm}, width=0.40\textwidth]{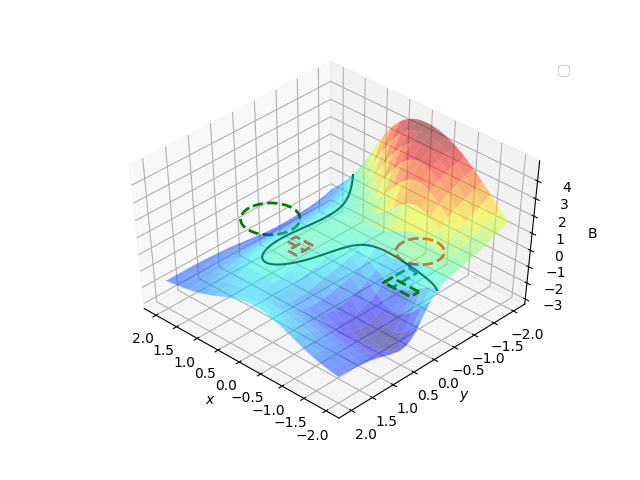}\hspace{0.5in}
\includegraphics[trim={1.5cm 1.5cm 1.5cm 1.5cm}, width=0.40\textwidth]{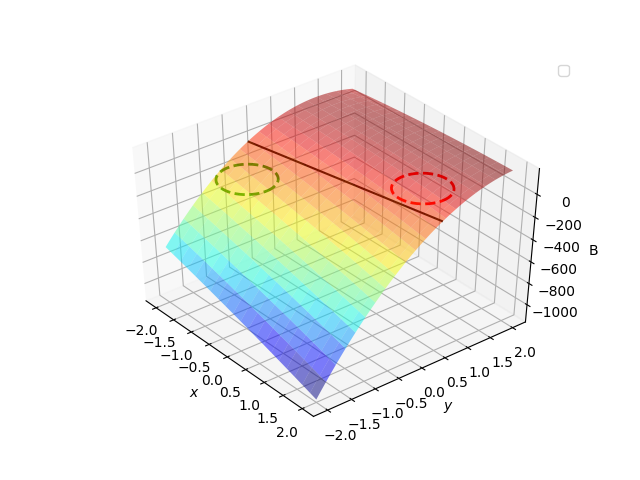}
\includegraphics[width=0.40\textwidth]{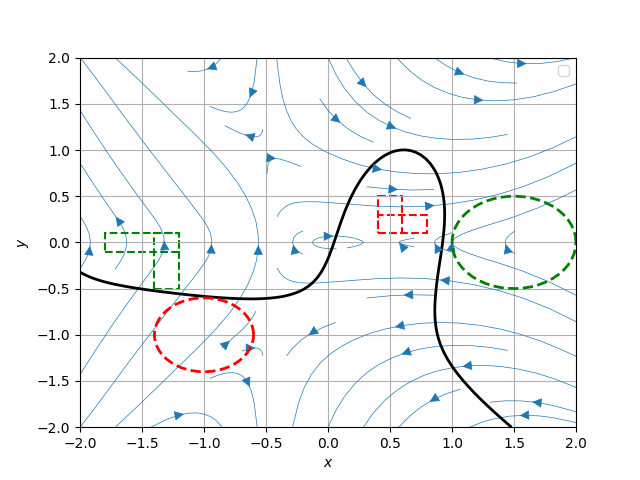}\hspace{0.5in}
\includegraphics[width=0.40\textwidth]{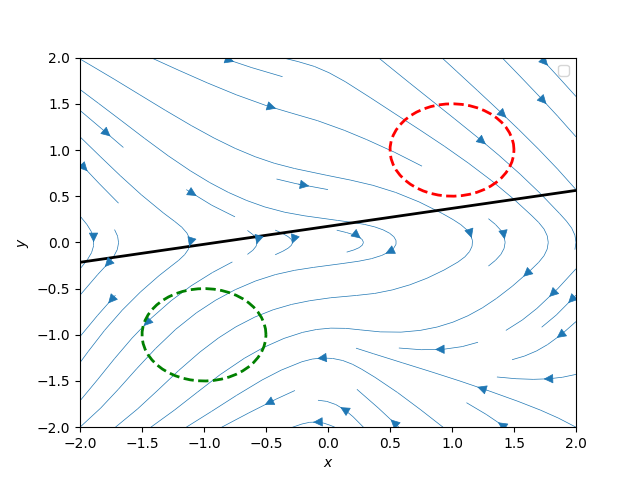}
\caption{The BC for the polynomial model (top left) and the hybrid model (top right) with the respective vector field (below).
}
\label{fig:poly-hybrid}
\end{figure}

\paragraph{\textbf{Polynomial Model} }
This model describes a polynomial system \cite{prajna2007framework} and presents initial and unsafe sets with complex, non convex shapes \cite{zhao2020barrier}, as follows: 
\begin{equation*}
\begin{cases}
\dot{x} = y, \\
\dot{y} = -x + \nicefrac{1}{3} \, x^3 - y, \text{ with domains }
\end{cases} 
\end{equation*}
\vspace{-0.2in}
\begin{align*}
& X = \{ -3.5 \leq x \leq 2,-2 \leq y \leq 1 \},  
\\
& X_0 = \{ (x-1.5)^2+y^2 \leq 0.25 \vee 
(x \geq -1.8 \wedge x \leq -1.2 \wedge y \geq -0.1 \wedge y \leq 0.1) 
\\
& \qquad\qquad \vee  (x \geq -1.4 \wedge x \leq -1.2 \wedge y\geq -0.5\wedge y \leq 0.1) \}, 
 \\
& X_u = \{ (x+1)^2+(y+1)^2 \leq 0.16 \vee
(x\geq 0.4 \wedge x\leq 0.6 \wedge y \geq 0.1 \wedge y\leq 0.5) 
\\
& \qquad\qquad \vee (x \geq 0.4 \wedge x \leq 0.8 \wedge y \geq 0.1 \wedge y \leq 0.3) \}. 
\end{align*}
Approaches from literature, such as (unsound) SOS-based procedures~\cite{legat2020sum,sloth2012compositional}, 
have required high-order polynomial templates, 
which has suggested the use of alternative activation functions. 
The BC, shown in Fig. \ref{fig:poly-hybrid}, is generated using a 10-layer NN with $\tanh$ activations. 
Needing just around 1 min and only 1000 initial samples, the overall procedure is 30 times faster than that in \cite{zhao2020barrier}. 

\paragraph{\textbf{Hybrid Model} }
We challenge our procedure with a 2-dimensional hybrid model, 
which extends beyond the capability of the results in \cite{zhao2020barrier}.  
This hybrid framework partitions the set $X$ into two non-overlapping subsets, $X_1$ and $X_2$. 
Each subset is associated to different model dynamics, respectively $f_1$ and $f_2$.  
In other words, the model trajectories evolve according to the $f_1$ dynamics when in $X_1$, and according to $f_2$ when in $X_2$. 
\begin{equation*}
f_1 = 
\begin{cases}
\dot{x} = y, \\
\dot{y} = - x - 0.5 x^3, \\
\end{cases}
\qquad
f_2 = 
\begin{cases}
\dot{x} = y, \\
\dot{y} =  x - 0.25 y^2, \\
\end{cases}   
\end{equation*}
\begin{equation*}
\text{with domain for } f_1 = \{ (x,y): x < 0 \}, \text{ domain for } f_2 = \{ (x,y): x\geq 0 \}, \text{ and sets } 
\end{equation*}
\vspace{-0.2in}
\begin{align*}
& X = \{ x^2 + y^2 \leq 4 \}, 
\qquad 
 X_0 = \{ (x+1)^2 + (y+1)^2 \leq 0.25 \}, 
 \\ 
& X_u = \{ (x-1)^2 + (y-1)^2 \leq 0.25 \}. 
\end{align*}
The structure of this model represents a non-trivial task for the verification engine, for which we employ the Z3 SMT solver: 
notice the dimensionally larger computation times.
The learning phase has instead been quite fast. 
The BC (Fig.\ref{fig:poly-hybrid}) is obtained from a single-layer NN comprising 3 neurons, using polynomial activations with $\gamma=2$, 
overall in less than 30 seconds, 
starting with an initial dataset of 500 samples. 

\paragraph{\textbf{Larger-dimensional Models} } 
We finally challenge our procedure with three high-order ODEs, respectively of order four, six and eight, 
to display the general applicability of our counter-example guided BC synthesis.   
We consider dynamical models described by the following differential equations: 
\begin{align}
& x^{(4)} 
+ 3980 x^{(3)} + 4180 x^{(2)} + 2400 x^{(1)} + 576 = 0,
\\
& x^{(6)} + 800 x^{(5)} + 2273 x^{(4)} 
+ 3980 x^{(3)} + 4180 x^{(2)} + 2400 x^{(1)} + 576 = 0, 
\\ \nonumber
& x^{(8)} + 20 x^{(7)} + 170 x^{(6)} + 800 x^{(5)} + 2273 x^{(4)} 
\\ 
& \qquad + 3980 x^{(3)} + 4180 x^{(2)} + 2400 x^{(1)} + 576 = 0, 
\end{align}
where we denote the $i$-th derivative of variable $x$ by $x^{(i)}$. 
We translate the ODE into a state-space model with variables $x_1, \dots, x_j$, where $j= \{4, 6, 8\}$, respectively. 
In all three instances, we select as spatial domain $X$ an hyper-sphere centred at the origin of radius 4; 
an initial set $X_0$ as hyper-sphere\footnote{We denote $\mathbf{1}^{[j]}$ the point of a $j$-dimensional state-space that has all its components equal to 1. For instance, $\mathbf{1}^{[3]}$ is the 3-dimensional point $[1, 1, 1]$. Similarly for $\mathbf{2}^{[j]}$.} centred at $\mathbf{+1}^{[j]}$ of radius 0.25; 
an unsafe set $X_u$ as an hyper-sphere centred at $\mathbf{-2}^{[j]}$ of radius 0.16. 
As an example, the eight-dimensional model presents the domains 
$ X = \{ x_1^2 + \ldots + x_8^2 \leq 4 \}$,  
$ X_0 = \{ (x_1-1)^2 + \ldots + (x_8-1)^2 \leq 0.25 \}$,  
$ X_u = \{ (x_1+2)^2 + \ldots + (x_8+2)^2 \leq 0.16 \}$. 
%
For the synthesis, we employ for all case studies a single-layer, 2-node architecture with polynomial ($\gamma$ = 1) activation function. 
Whilst in particular the verification engine is challenged from the high dimensionality of the models, 
the CEGIS procedure returns a valid barrier certificate in only two iterations and with very reasonable run times.  



\paragraph{\textbf{Experimental Repeatability and Codebase Robustness}}

Our algorithm uses pseudo-random number generators in two
instances: 
the initial samples set  
and the generation of counterexamples. 
To test the robustness of the overall algorithm, 
we now outline a statistical analysis of repeated tests, 
where the results are obtained after a single run of the algorithm. 
We report learning time, verification time and number of iterations, 
over 100 runs, 
in Table \ref{tab:100-results}. 
The obtained statistics include the average running times and the average number of iterations, 
along with the minimum and maximum values over the 100 runs. 
Note that these values represent the average learning and verification times over single iterations of the CEGIS procedure: 
that is, unlike the cumulative results in Table \ref{tab:results}, these represent the average runtimes of {\it single calls} of the learning and verification engines, respectively.  
In the section above we have instead reported the {\it sum} of the learning and verification run times, over the overall number of iterations of the CEGIS procedure. 
Table \ref{tab:100-results} shows learning and verification results that are similar to the ones reported in Table \ref{tab:results}, consistently finding barrier certificates for every benchmark: this clearly indicates the robustness of our approach and the reliability of our codebase.

\begin{table}[]
\centering
\begin{tabular}{c|c|c|c}
Benchmark & Learner & Verifier & Iters \\ \hline
Darboux Model & 27.47 [26.9, 27.9] & 0.024 [0.021, 0.028] & 1 [1, 1] \\
Exponential Model & 12.51 [8.1, 21.5] & 0.002 [0.001, 0.003] & 2 [2, 2] \\\
Obstacle Avoidance & 22.07 [17.9, 27.5] & 0.005 [0.004, 0.007] & 1 [1, 1]\\
Polynomial Model & 29.63 [26.5, 33.8] & 0.01 [0.006, 0.017] & 2 [2, 2]\\
Hybrid Model & 0.55 [0.5, 0.8] & 2.50 [2.0, 2.9] & 1.9 [1, 10]
\\
4-d ODE Model & 36.46  [35.25 39.44] & 0.66 [0.32, 1.13] & 2 [2,2] 
\\
6-d ODE Model & 36.53  [35.06, 37.83] & 0.67  [0.33, 1.12] & 2  [2, 2] 
\\
8-d ODE Model & 41.94 [35.76, 50.80] & 32.06 [9.02, 52.05] & 2  [2, 2]
\end{tabular}
\vspace{1pt}
\caption{Average learning and verification run times (in seconds), 
and average number of iterations of the CEGIS procedure, over 100 runs.  
The square brackets contain the minimum and maximum values obtained.}
\label{tab:100-results}
\end{table}


\section{Conclusions and Future Work}
\label{sec:conlusion}

We have presented a new inductive, formal, automated technique to synthesise neural-based barrier certificates for polynomial and non-polynomial, continuous and hybrid dynamical models.  
Thanks to a number of architectural choices for the new procedure, 
our method requires less training data and thus displays faster learning, 
as well as quicker verification time, 
than state-of-the-art techniques. 

Towards increased automation, future work includes the development of an automated selection of activation functions that are tailored to the dynamical models of interest. 


\bibliographystyle{plain}
\bibliography{neuralbc_biblio}

\end{document}

%% file: Figures/cegis.tex
\usetikzlibrary{arrows,positioning} 
\tikzset{
    >=stealth',
    punkt/.style={
           rectangle,
           rounded corners,
           draw=black, very thick,
           text width=12em,
           minimum height=6em,
           text centered},
    pil/.style={
           ->,
           thick,
           shorten <=2pt,
           shorten >=2pt,}
}

\begin{tikzpicture}
    \node[shape=rectangle,rounded corners, draw, minimum width=1.7cm, minimum height=1cm, align=center] (lea) at (0,0) {Learner \\ NN};
    \node[shape=rectangle, rounded corners, draw, minimum width=1.7cm, minimum height=1cm, align=center] (ver) at (3.5,0) {Verifier \\ SMT};
    
    \draw[pil, bend left] (ver) edge node[below]{counter-example} (lea);
	\draw[pil, bend left] (lea) edge node[above]{candidate $BC$} (ver);

    \node (out) at (6.0,0) {$B$};
    \draw[pil] (ver) edge node[above]{valid} (out);

\end{tikzpicture}

%% file: Figures/algorithm.tex
\begin{algorithm}
\caption{Synthesis of Neural Barrier Certificate}
\begin{multicols}{2}
\begin{algorithmic}
\Function{Learner}{$S$, $f$}
\Repeat
\State $B(S)$ $\leftarrow$ NN($S$)
\State $\dot{B}(S)$ $\leftarrow$ $\nabla B(S) \cdot f(S)$ 
\State compute loss $L$, update NN
\Until{convergence} 
\State \Return NN
\EndFunction 
\\
\Function{Verifier}{$B$, $\dot{B}$}
\State encode conditions in \eqref{eq:verifier-conditions}
\State Cex or {\tt unsat} $\leftarrow$ SMTcheck($B$, $\dot{B}$) 
\State \Return Cex or {\tt unsat}
\EndFunction \\ 
\Function{CEGIS}{$f$}
\State initialise NN, $S$
\Repeat
\State NN $\leftarrow$ \Call{Learner}{$S$, $f$}
\State $B(x)$, $\dot{B}(x)$ $\leftarrow$ Translate(NN, $f$)
\State Cex or {\tt unsat} $\leftarrow$ \Call{Verifier}{$B$, $\dot{B}$}
\State S $\leftarrow$ S $\cup$ Cex
\Until{{\tt unsat}}
\State \Return $B(x)$, $\dot{B}(x)$
\EndFunction
\State
\State 
\\
\end{algorithmic}
\end{multicols}
\end{algorithm}